\documentclass{nature}
\usepackage[dvips]{graphicx}

\def\kms{\hbox{km s$^{-1}$}}
\def\VLSR{\hbox{$V_{\rm LSR}$}}

\def\sun{\hbox{$\odot$}}

\def\lesssim{\mathrel{\hbox{\rlap{\hbox{\lower4pt\hbox{$\sim$}}}\hbox{$<$}}}}
\def\gtrsim{\mathrel{\hbox{\rlap{\hbox{\lower4pt\hbox{$\sim$}}}\hbox{$>$}}}}

\def\arcdeg{\hbox{$^\circ$}}

\def\arcsec{\hbox{$^{\prime\prime}$}}

\title{Millimetre-wave Emission from an Intermediate-Mass Black Hole Candidate in the Milky Way}

\author{Tomoharu Oka$^{1,2}$, Shiho Tsujimoto$^2$, Yuhei Iwata$^2$, Mariko Nomura$^1$, \&\ Shunya Takekawa$^2$}

\begin{document}

\maketitle

\begin{affiliations}
\item Department of Physics, Institute of Science and Technology, Keio University, 3-14-1 Hiyoshi, Kohoku-ku, Yokohama, Kanagawa 223-8522, Japan
\item School of Fundamental Science and Technology, Graduate School of Science and Technology, Keio University, 3-14-1 Hiyoshi, Kohoku-ku, Yokohama, Kanagawa 223-8522, Japan
\end{affiliations}

\begin{abstract}
It is widely accepted that black holes (BHs) with masses greater than a million solar masses ($M_{\sun}$) lurk at the centres of massive galaxies. The origins of such `supermassive' black holes (SMBHs) remain unknown\cite{Djorgovski99}, while those of stellar-mass BHs are well-understood. One possible scenario is that intermediate-mass black holes (IMBHs), which are formed by the runaway coalescence of stars in young compact star clusters\cite{Portegies99}, merge at the centre of a galaxy to form an SMBH\cite{Ebisuzaki01}. Although many candidates for IMBHs have been proposed to date, none of them are accepted as definitive. Recently we discovered a peculiar molecular cloud, CO--0.40--0.22, with an extremely broad velocity width near the centre of our Milky Way galaxy. Based on the careful analysis of gas kinematics, we concluded that a compact object with a mass of $\sim\!10^5$ $M_{\sun}$ is lurking in this cloud\cite{Oka16}. Here we report the detection of a point-like continuum source as well as a compact gas clump near the center of CO--0.40--0.22. This point-like continuum source (CO--0.40--0.22$^*$) has a wide-band spectrum consistent with 1/500 of the Galactic SMBH (Sgr A$^*$) in luminosity. Numerical simulations around a point-like massive object reproduce the kinematics of dense molecular gas well, which suggests that CO--0.40--0.22$^*$ is the most promising candidate for an intermediate-mass black hole.
\end{abstract}


CO--0.40--0.22 is a compact cloud ($\sim\!5$ pc) with an extremely broad velocity width ($\sim\!100$ \kms) and very high CO {\it J}=3--2/{\it J}=1--0 intensity ratio ($\geq \!1.5$) detected at a projected distance of $\sim\!60$ pc away from the galactic nucleus\cite{Oka12}. It belongs to a peculiar category of molecular clouds called high-velocity compact clouds (HVCCs) that were originally identified in the CO {\it J}=1--0 survey data\cite{Oka98, Oka99, Oka01}. 
CO--0.40--0.22 is only a dense cloud with a negative velocity detected in HCN {\it J}=4--3 line within the $0.3\arcdeg\!\times\! 0.3\arcdeg$ area including it\cite{Oka16}. It has a continuous and roughly straight entity in the position-velocity maps, seeming not to be an aggregate of unrelated clouds with less broad velocity widths.  
The kinematical structure of CO--0.40--0.22 can be explained as being due to a gravitational kick experienced by the molecular cloud caused by an invisible compact object with a mass of $\sim\!10^5$ $M_{\sun}$. The compactness and absence of a counterpart at other wavelengths suggest that this massive object is an inactive IMBH, which is not currently accreting matter. This is the second-largest black hole candidate in the Milky Way galaxy affter Sgr A$^*$, as well as the second IMBH candidate in the Galaxy after that in the nuclear subcluster IRS13E ($M_{\rm BH}\!\simeq\!1300$ $M_{\sun}$)\cite{Maillard04}$^{,}$\cite{Schoedel05}.

ALMA band 6 observations towards CO--0.40--0.22 have provided high-resolution HCN {\it J}=3--2 (265.9 GHz) and CO {\it J}=2--1 (230.5 GHz) images. Dense molecular gas traced by HCN {\it J}=3--2 emission seems to concentrate near the centre of CO--0.40--0.22 as previously determined by the coarse-resolution ASTE HCN {\it J}=4--3 map (Fig. 1a). The displacement of $0.2$ pc from the centre is within the ASTE beamwidth ($22^{\prime\prime}\!=\!0.9$ pc). This dense gas clump is very compact ($\sim\! 0.3$ pc) and has a broad velocity width ($\sim\! 100$ \kms). 
We also see $\sim\! 20$ rather faint clumps. These faint clumps and the central dense clump occupy $\sim\!10$ \%\ of the field of view. The faint clumps have velocity widths less than 20 \kms. Thus the chance probability of their alignment over $\sim\! 100$ \kms\ width is less than $(0.1/(100/20))^{100/20}\!=\! 10^{-8.5}$.  
The main body of the central clump appears at $\VLSR\!=\!-80$ to $-40$ \kms, being associated with high-velocity components which reach $\VLSR\!=\!-105$ and $-5$ \kms\ (Fig. 2). It is slightly elongated in roughly the same direction as the major axis of CO--0.40--0.22 with a steep velocity gradient from southeast to northwest. The mass of the clump was estimated from the HCN {\it J}=3--2 integrated intensity to be $40$ $M_{\sun}$ according to an excitation model at $T_{\rm k}\!=\!60$ K and $n({\rm H}_2)\!=\!10^{6.5}$ cm$^{-3}$. On the other hand, the size parameter $S\!=\!0.15$ pc and velocity dispersion $\sigma_{\rm V}\!=\!22$ \kms\ give a virial theorem mass of $M_{\rm vir}\!\simeq\!4.1\!\times\! 10^3$ $M_{\sun}$, which indicates that the clump must not be bound by its self-gravity. 

In the near-vicinity of the compact dense gas clump, we detected a point-like continuum source (Fig. 1b). It is located at $(l, b)\!=\!(-0.3983\arcdeg, -0.2235\arcdeg)$ in galactic coordinates. Its size and shape coincide with those of the primary beam of our ALMA observations ($1.35\arcsec\!\times\!0.55\arcsec@266$ GHz). This point-like source (CO--0.40--0.22$^*$) is prominent in both of 231 and 266 GHz images precisely at the same position with flux densities of $8.38\!\pm\!0.34$ and $9.91\!\pm\!0.74$ mJy, respectively. These flux densities provide a millimetre-wave luminosity of $\nu L_{\nu}\!\sim\! 10^{32.3}$ erg s$^{-1}$ at a distance of 8.3 kpc\cite{Gillessen09}. This CO--0.40--0.22$^*$ was the only significant source within the $80\arcsec$ diameter field of view.

The spectral index at 231/266 GHz is $\alpha\!=\!1.18\!\pm\!0.65$, which indicates an inverted non-thermal spectrum or a thermal spectrum near the peak. A least-squares fit to the 231 and 266 GHz fluxes by the Planck function gives $T\!\simeq\!9$ K. This temperature is far lower than dust temperatures in protoplanetary disks and even lower than the typical dust temperature in the central molecular zone of our Galaxy ($T_{\rm d}=21\pm 2$ K)\cite{Pierce00}. Alternatively, a background submillimetre galaxy at high redshift may show a millimetre-wave spectrum similar to CO--0.40--0.22$^*$. However, the chance probability of a submillimetre galaxy with a flux greater than 10 mJy within a $10\arcsec$ diameter circle is estimated to be only $\sim 0.1\,\%$ based on the observed number density of such objects\cite{Hatsukade11}. Thus, the thermal origin of millimetre-wave emission from CO--0.40--0.22$^*$ is implausible. Rather, it may be an inverted non-thermal (power-law) spectrum like that of Sgr A$^*$. The inverted spectrum could be due to the absorption. The theoretically steepest synchrotron self-absorbed spectrum has $\alpha\!=\!5/2\;$\cite{Rybicki86}.  

What is CO--0.40--0.22$^*$? The observed flux densities correspond to $\sim\!1/500$ of the Sgr A$^*$ luminosity at the same frequency. By inspecting the X-ray data obtained by XMM Newton, we obtained an upper limit to the $1\mbox{--}7$ keV flux density of CO--0.40--0.22$^*$: $1.4\!\times\!10^{-14}$ erg cm$^{-2}$ s$^{-1}$ ($\sim\!1$ nJy). This is also consistent with the 1:500 scale spectrum of Sgr A$^*$ (Fig.3). All of the observed facts indicate it is possible that millimetre-wave emission from CO--0.40--0.22$^*$ are presumably from the near-vicinity of an IMBH.

Now, we have the accurate position of the IMBH candidate. We accordingly refined the gravitational kick model for CO--0.40--0.22. We basically followed the parameters employed in the previous paper\cite{Oka16} except for the inclination of the orbital plane ($i\!=\!70\arcdeg$ with respect to the line of sight). In the new simulation, we included self-gravity within the model cloud and set it initially virialized. We placed a point-like mass of $10^5\,M_{\sun}$ at the centre and a spherical cloud 
at $(X, Y)\!=\!(9.8\,\mbox{pc}, -0.65\,\mbox{pc})$ in the coordinates defined by the orbital plane (see Fig. 4 for the configuration). The origin was set to the centre of gravity. Then, the model cloud was thrown with an initial velocity of $(v_{\rm X}, v_{\rm Y})\!=\!(-8.19\,\kms , 0.4\,\kms)$. The simulated cloud just after the pericentre passage reproduced the bulk kinematics of CO--0.40--0.22 well. It also reproduced the spatial-velocity behaviour of the compact dense gas clump very well (Figs. 1b, 2).

According to the refined gravitational kick scenario, CO--0.40--0.22$^*$ must be as massive as $10^5$ $M_{\sun}$. Its radius must be significantly smaller than 0.022 pc, which corresponds to the minor radius of the primary beam. These mass and size values gives a lower limit to the average mass density of $\rho\!\simeq\! 10^{9.4}$ $M_{\sun} \mbox{pc}^{-3}$. This mass density is two orders of magnitude greater than that of the core of M15, which is one of the most densely packed (core-collapsed) globular clusters in the Milky Way galaxy\cite{Djorgovski84}. If CO--0.40--0.22$^*$ is a $10^5$ $M_{\sun}$ globular cluster, it must have a luminosity of $\sim\!10^5\,L_{\sun}$\cite{McLaughlin2000}. Non-detection of an infrared counterpart disfavours the stellar cluster interpretation, unless the cluster consists almost entirely of dark stellar remnants, i.e. neutron stars and BHs. Such a massive clustering of dark stellar remnants is implausible. Therefore, CO--0.40--0.22$^*$ is most likely a radio source closely associated with a single $10^5$ $M_{\sun}$ BH.

Where did the IMBH come from? Runaway stellar mergers at centres of dense star clusters can produce massive BHs. However, the ratio of the BH mass ($M_{\rm BH}$) to the parent cluster mass ($M_{\rm C}$) is typically $\sim\!1/1000$\cite{Marchant80, Portegies99}, which is also between the masses of an SMBH and the galactic bulge\cite{Marconi03}. According to this $M_{\rm BH}$-$M_{\rm C}$ relation, a $10^5$ $M_{\sun}$ BH requires a parent cluster of $\sim\!10^8$ $M_{\sun}$, which is rather categorized in the upper hierarchy, i.e. a dwarf galaxy. Black holes with masses of less than a million $M_{\sun}$ have been suggested in the nuclei of nearby dwarf galaxies\cite{Moran14}. It is believed that large galaxies such as the Milky Way grew to their present form by cannibalizing their smaller neighbours. In fact, a minor merger event $\sim\!200$ Myr ago was suggested by infrared stellar populations in the central kiloparsec\cite{vanLoon03}. To date, more than 50 satellite galaxies have been discovered within $420$ kpc from the Milky Way. We suggest that CO--0.40--0.22$^*$ used to be the nucleus of a dwarf galaxy which was cannibalized by our Milky Way.

The compact radio source in HVCC CO--0.40--0.22 strongly supports our theory that a gravitational kick by an IMBH is responsible for the extraordinarily broad velocity width of this HVCC. The extreme dimness of CO--0.40--0.22$^*$ may inspire the development of new radiatively inefficient accretion flow (RIAF) models. Future multi-wavelength observations and flux variation measurements of CO--0.40--0.22$^*$ will elaborate on such RIAF models with an isolated massive BH. They may possibly show phenomena unique to a BH, such as quasi-periodic oscillations. The case of CO--0.40--0.22 opens a new perspective on the study of other HVCCs, as well as compact high-velocity features in the Galactic disk. Some of them may contain nonluminous BHs. For example, an ultra-high-velocity gas in the W44 supernova remnant\cite{Sashida13} is suggested to have been accelerated by the explosion or irruption of a BH\cite{Yamada17}. We recently deteceted two ultracompact high-velocity features in the vicinity of the Galactic circumnuclear disk, which also could be driven by the plunge of inactive BHs\cite{Takekawa17}. Theoretical studies have predicted that 100 million to 1 billion BHs should exist in the Milky Way\cite{Agol02}, although only 60 or so have been identified through observations to date\cite{Corral16}. Further detection of such compact high-velocity features in various environments may increase the number of nonluminous BH candidate and thereby increase targets to search for evidential proof of general relativity. It will make a great contribution to the progressive development of modern physics.


\begin{thebibliography}{1}
\bibitem{Djorgovski99} Djorgovski, S. G., Volonteri, M., Springel, V., Bromm, V., \&\ Meylan, G. in {\it The Eleventh Marcel Grossmann Meeting on Recent Developments in Theoretical and Experimental General Relativity, Gravitation and Relativistic Field Theories} (eds Kleinert, H., Jantzen R. T., \&\ Ruffini, R.) 340--367 (World Scientific, 2008)
\bibitem{Portegies99} Portegies Zwart, S. F., Makino, J., McMillan, S. L. W. \&\ Hut, P. Star cluster ecology. III. Runaway collisions in young compact star clusters. {\it Astron. Astrophys.} {\bf 348}, 117--126 (1999).
\bibitem{Ebisuzaki01} Ebisuzaki, T. {\it et al.} Missing link found? The ``runaway" path to supermassive black holes. {\it Astrophys. J.} {\bf 562}, L19--L22 (2001).
\bibitem{Oka16} Oka, T., Mizuno, R., Miura, K., \&\ Takekawa, S. Signature of an intermediate-mass black hole in the central molecular zone of our galaxy. {\it Astrophys. J.} {\bf 816}, L7 (2016).
\bibitem{Oka12} Oka, T. {\it et al.} ASTE CO {\it J}=3--2 survey of the Galactic Center. {\it Astrophys. J. Suppl.} {\bf 201}, 14--25 (2012).
\bibitem{Oka98} Oka, T., Hasegawa, T., Sato, F., Tsuboi, M. \&\ Miyazaki, A. A large-scale CO survey of the Galactic Center. {\it Astrophys. J. Suppl.} {\bf 118}, 455--515 (1998).
\bibitem{Oka99} Oka, T. {\it et al.} A high-velocity molecular cloud near the center of the Galaxy. {\it Astrophys. J.} {\bf 515}, 249--255 (1999).
\bibitem{Oka01} Oka, T., Hasegawa, T., Sato, F., Tsuboi, M. \&\ Miyazaki, A. A hyperenergetic CO shell in the Galactic Center molecular cloud complex. {\it Publ. Astron. Soc. Japan} {\bf 53}, 787--791 (2001).
\bibitem{Maillard04} Maillard, J. P., Paumard, T., Stolovy, S. R., \&\ Rigaut, F. The nature of the Galactic Center source IRS 13 revealed by high spatial resolution in the infrared. {\it Astron. Astrophys.} {\bf 423}, 155--167 (2004).
\bibitem{Schoedel05} Sch\"odel, R., Eckart, A., Iserlohe, C., Genzel, R., \&\ Ott, T.  Black Hole in the Galactic Center Complex IRS 13E? {\it Astrophys. J.} {\bf 625}, L111--L114 (2005).
\bibitem{Gillessen09} Gillessen, S. {\it et al.} Monitoring stellar orbits around the massive black hole in the Galactic Center. {\it Astrophys. J.} {\bf 692}, 1075--1109 (2009).
\bibitem{Pierce00} Pierce-Price, D. {\it et al}. A deep submillimeter survey of the Galactic Center. {\it Astrophys. J.} {\bf 545}, L121--L125 (2000).
\bibitem{Hatsukade11} Hatsukade, B., {\it et al.} AzTEC/ASTE 1.1-mm survey of the AKARI Deep Field South: source catalogue and number counts. {\it Mon. Not. R. Astron. Soc.} {\bf 411}, 102--116 (2011).
\bibitem{Rybicki86} Rybicki G. B., \&\ Lightman A. P. Radiative Processes in Astrophysics, Wiley-VCH  (1986).
\bibitem{Djorgovski84} Djorgovski, S. \&\ King, I. R. Surface photometry in cores of globular clusters. {\it Astrophys. J.} {\bf 277}, L49--L52 (1984).
\bibitem{McLaughlin2000} McLaughlin, D. E., Binding energy and the fundamental plane of globular clusters. {\it Astrophys. J.} {\bf 539}, 618--640 (2000).
\bibitem{Marchant80} Marchant, A. B. \&\ Shapiro, S. L. Star clusters containing massive, central black holes. III -- Evolution calculations. {\it Astrophys. J.} {\bf 239}, 685--704 (1980).
\bibitem{Marconi03} Marconi, A., \&\ Hunt, L. K. The relation between black hole mass, bulge mass, and near-infrared luminosity. {\it Astrophys. J.} {\bf589}, L21--L24 (2003).
\bibitem{Moran14} Moran, E. C., {\it et al.} Black holes in the centers of nearby dwarf galaxies. {\it Astron. J.} {\bf 148}, 136--157 (2014).
\bibitem{vanLoon03} van Loon, J. Th. {\it et al.} Infrared stellar populations in the central parts of the Milky Way galaxy. {\it Astron. Astrophys.} {\bf 338}, 857--879 (2003).
\bibitem{Sashida13} Sashida, T. {\it et al.} Kinematics of shocked molecular gas adjacent to the supernova remnant W44. {\it Astrophys. J.} {\bf 774}, 10--16 (2013).
\bibitem{Yamada17} Yamada, M., {\it et al.} Kinematics of ultra-high-velocity gas in the expanding molecular shell adjacent to the W44 supernova remnant. {\it Astrophys. J.} {\bf 834}, L3 (2017).
\bibitem{Takekawa17} Takekawa, S., Oka, T., Iwata, Y., Tokuyama, S., \&\ Nomura, M. Discovery of two small high-velocity compact clouds in the central 10 parsecs of the Galaxy. {\it Astrophys. J.}  in press (2017). 
\bibitem{Agol02} Agol, E. Kamionkowski, M., Koopmans, L\'eon V. E.\&\ Blandford, Roger D. Finding black holes with microlensing. {\it Astrophys. J.} {\bf 576}, L131--L135 (2002). 
\bibitem{Corral16} Corral-Santana, J. M., {\it et al.} BlackCAT: A catalogue of stellar-mass black holes in X-ray transients. {\it Astron. Astrophys.} {\bf 587}, A61 (2016).
\end{thebibliography}

\clearpage
\begin{addendum}
\item This paper makes use of the following ALMA data: ADS/JAO.ALMA\#2012.1.00940.S. ALMA is a partnership of ESO (representing its member states), NSF (USA) and NINS (Japan), together with NRC (Canada), MOST and ASIAA (Taiwan), and KASI (Republic of Korea), in cooperation with the Republic of Chile. The Joint ALMA Observatory is operated by ESO, AUI/NRAO and NAOJ. We thank the ALMA staff for the operation of the array and delivering the qualified data. We also thank S. Nakashima and M. Nobukawa for calculating the upper limit to the X-ray flux, and A. E. Higuchi for helping in ALMA data reduction with CASA. T.O. acknowledges support from JSPS Grant-in-Aid for Scientific Research (B) No. 15H03643.
\end{addendum}
\clearpage

{\Large\bfseries\noindent\sloppy\textsf{Author Contributions}}
\newline
T.O. directed the research, analysed the data, and wrote the manuscript. S.T. and M.N. performed the model calculation. Y.I. and S.T. contributed to the analyses and discussion.

{\Large\bfseries\noindent\sloppy\textsf{Author Information}}
\newline
Reprints and permission information are available at www.nature.com/reprints. The authors declare no competing financial interests. Correspondence and requests for materials should be addressed to T.O. (tomo@phys.keio.ac.jp).
\clearpage

{\Large\bfseries\noindent\sloppy\textsf{Figure Legends}}

\begin{figure}[htbp]
\begin{center}
\includegraphics[width=0.98\textwidth]{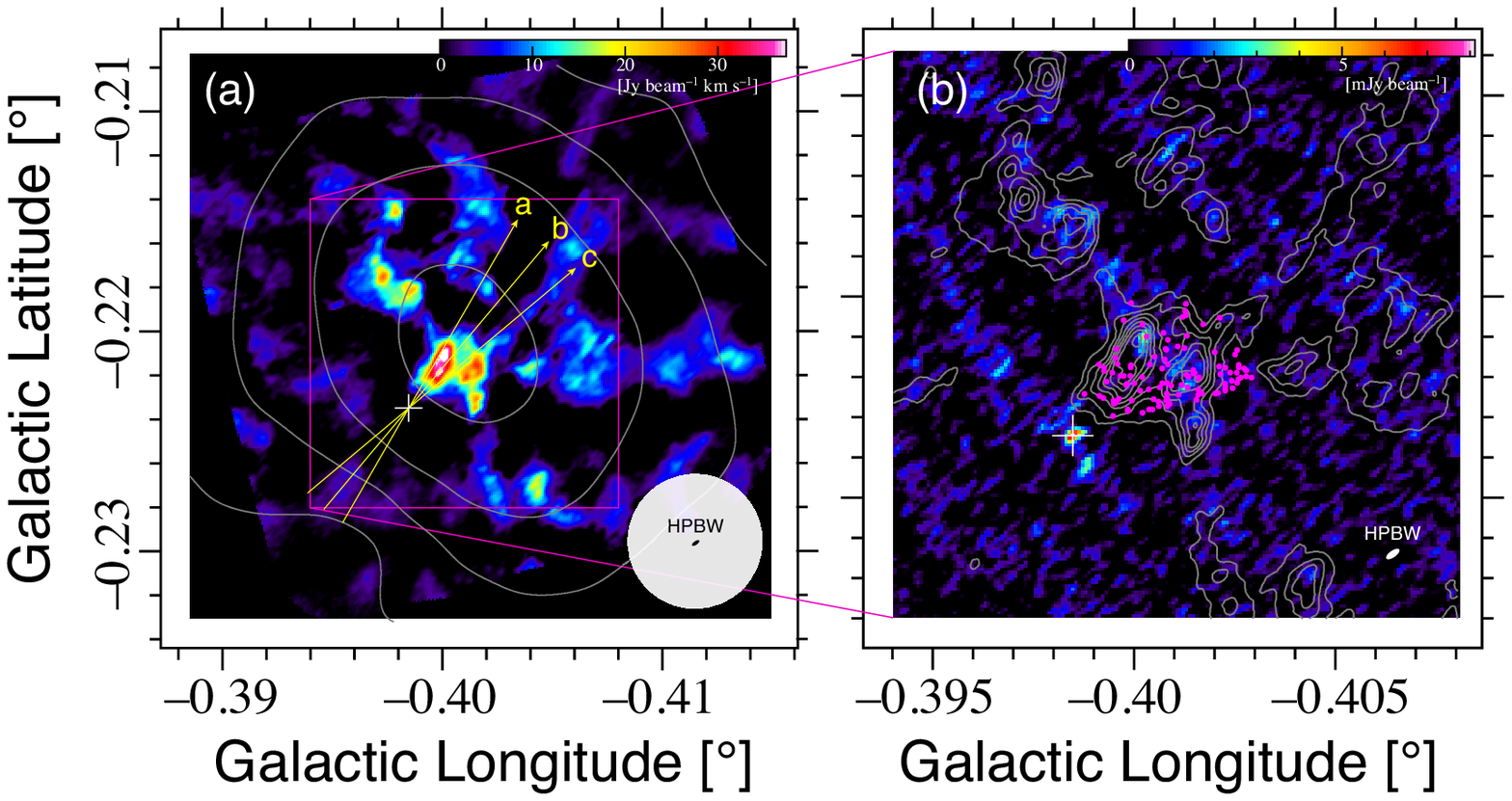}
\end{center}
\caption{
\noindent
{\bf ALMA views of CO--0.40--0.22.} {\bf a)} Colour map of HCN {\it J}=3--2 emission integrated over $\VLSR\!=\!-110\mbox{--}0$ \kms. The white contours show the same map of HCN {\it J}=4--3 emission obtained with the ASTE 10 m telescope\cite{Oka16}. The contour levels are 50, 100, 200, and 400 K \kms .  Beamsizes of ASTE and ALMA are also presented by a white filled circle and black filled ellipse, respectively. Yellow arrows indicate the lines along which position-velocity slices are made (Fig. 2), and the white cross shows the location of CO--0.40--0.22$^*$. {\bf b)} Zoom-in images of HCN {\it J}=3--2 velocity-integrated emission (contours) and 266 GHz continuum emission (colour). The contour interval is 5 Jy beam$^{-1}$ \kms . The white filled ellipse shows the ALMA beamsize. Magenta dots show the loci of cloud particles in the gravitational kick model at $t\!=\!7.2\times 10^5$ yr.
}
\label{fig1}
\end{figure}

\clearpage
\begin{figure}[htbp]
\begin{center}
\includegraphics[width=0.9\textwidth]{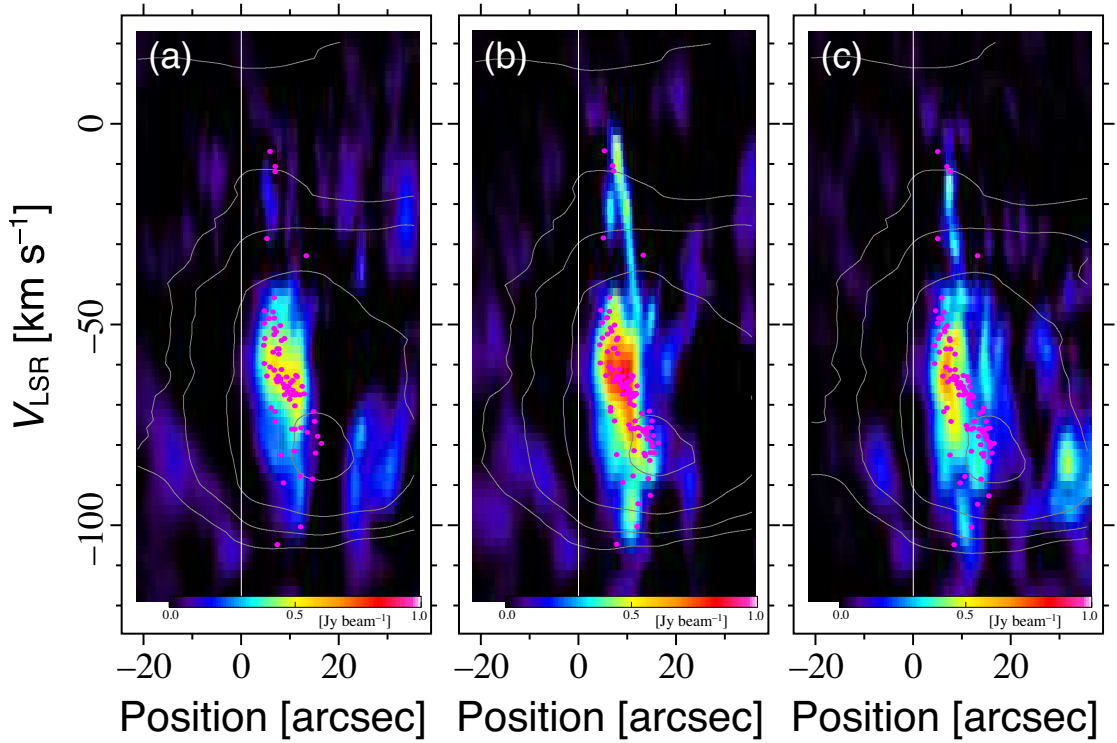}
\end{center}
\caption{
\noindent
{\bf Gas kinematics around CO--0.40--0.22$^*$.}
Position-velocity maps along the yellow arrows in Fig. 1{\bf a} labelled as {\bf a}, {\bf b}, and {\bf c}, respectively. Magenta dots show the loci of cloud particles in the gravitational kick model projected to each position-velocity plane. The white contours show the same maps of HCN {\it J}=4--3 emission obtained with the ASTE. The contour levels are 2, 4, 8, and 16 K. The vertical solid line in each panel describes the position of CO--0.40--0.22$^*$.  
}
\label{fig2}
\end{figure}

\clearpage
\begin{figure}[htbp]
\begin{center}
\includegraphics[width=0.7\textwidth]{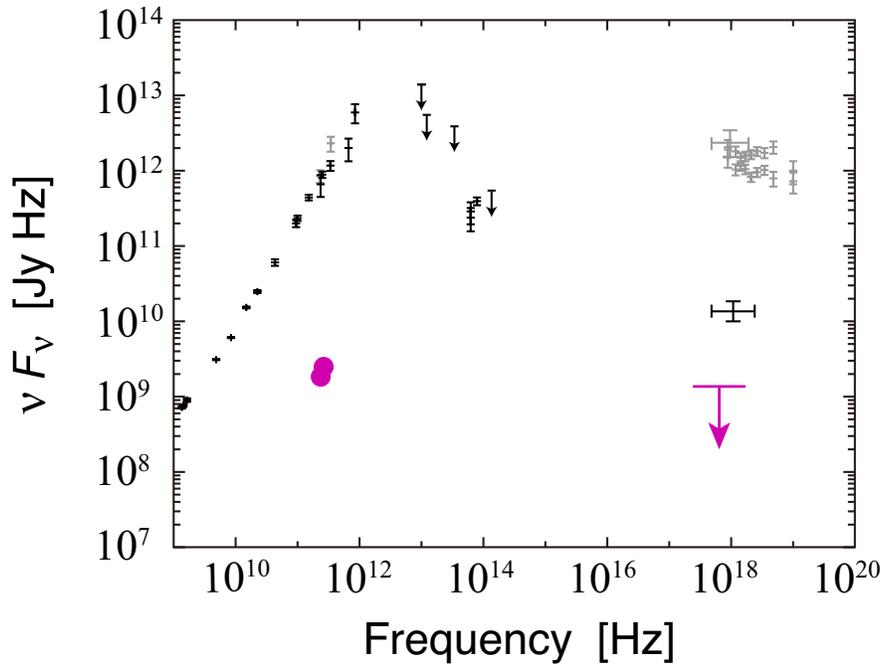}
\end{center}
\caption{
\noindent
{\bf Wide-band spectrum of CO--0.40--0.22$^*$.}
Wide-band spectrum of Sgr A$^*$ in the quiescent (black) and flare-up (grey) states. Error bars are $1\,\sigma$ uncertainties. The magenta filled circles show the 231 and 266 GHz fluxes of CO--0.40--0.22$^*$. Uncertainties are smaller than the size of the symbol. The magenta bar with a downward arrow denotes the $3\,\sigma$ upper limit to the X-ray flux.
}
\label{fig3}
\end{figure}

\clearpage
\begin{figure}[htbp]
\begin{center}
\includegraphics[width=0.7\textwidth]{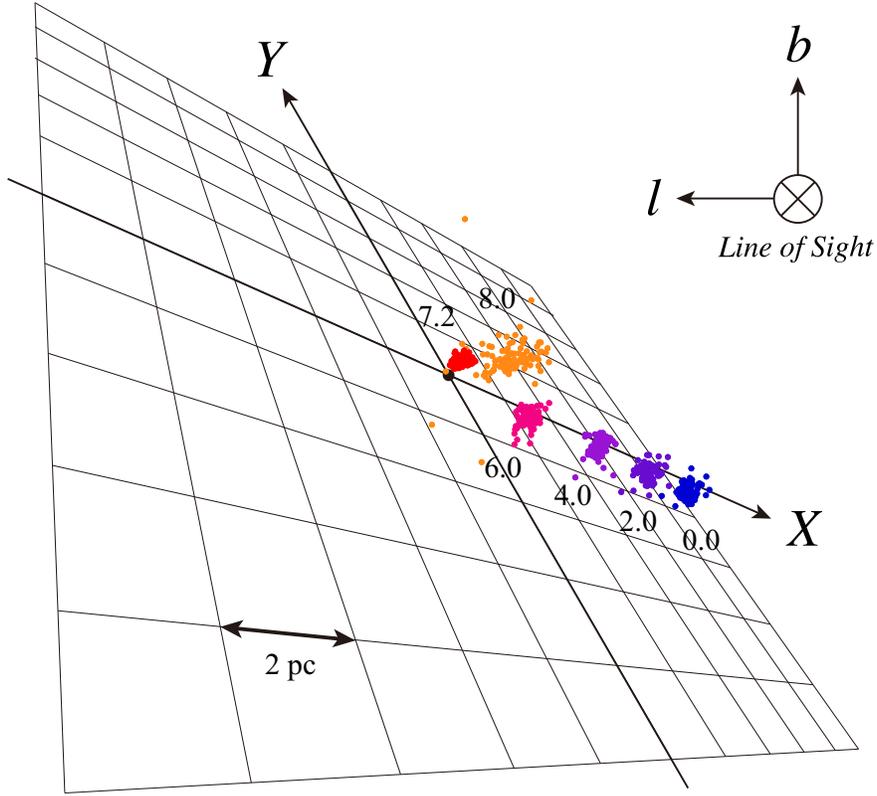}
\end{center}
\caption{
\noindent
{\bf N-body simulations of the gravitational kick.}
Geometric configuration of the model cloud and a $10^5\,M_{\sun}$ point-like mass drawn in perspective from a distance of 30 pc. The horizontal direction is the Galactic longitude ($l$) and the vertical direction is the Galactic latitude ($b$). The orbital plane is displayed by XY axes. The loci of cloud particles are illustrated at five time slices: $t\!=\!0$, $2$, $4$, $6$, $7.2$, and $8.0$ (unit: $10^5$ yr). One hundred five particles that overlap with the HCN {\it J}=3--2 clump in the plane of the sky at $t\!=\!7.2\times 10^5$ yr (Fig.1b) are traced.
}
\label{fig4}
\end{figure}

\clearpage

\begin{methods}
\subsection{ALMA observations and data reduction.}
Our cycle 1 ALMA observations (ALMA/2012.1.00940.S) in Band 6 were taken on 2013 July 8, October 6--7, 2014 April 13, May 11, December 29, and 2015 January 22. The 230.5 GHz (CO {\it J}=2--1) data were obtained with seven 7-m antennas and 39 12-m antennas, while the 265.9 GHz (HCN {\it J}=3--2) data were obtained with eight or ten 7-m antennas and 30 or 31 12-m antennas. For CO--0.40--0.22 observations, we configured the correlators to the 937.5 MHz bandwidth (3840 channels) mode. Each target line was centered in one spectral window. Amplitude calibration was based on Mars, Titan, Neptune, and J2232+117. We used J1924--2914, J1517--2422, J2427--4206, J1733--1304, J2258--2758, and J1517--2422 as bandpass calibrators.  J1744--3116, J2337--0230, and J1745--2900 were used as phase calibrators. Calibrator selections depend on epochs.  

We reduced the data with Common Astronomy Software Applications (CASA) version 4.4.0-REL with the calibration script supplied by the East Asian ALMA Regional Centre (EA ARC). The data were splitted into line and continuum data sets by using the task {\it uvcontsub()}. For spectral line data, we reduced the data with channels from $-150$ \kms\ to $+50$ \kms\ with a $2$ \kms\ width and a cell size of $0.2\arcsec$.  We ran the cleaning algorithm with natural weighting. The angular sizes of the cleaned beam full-width at half-maximum are $1.87\arcsec\!\times\! 1.14\arcsec$ and $1.52\arcsec\!\times\! 0.60\arcsec$, at 230.5 GHz and 265.9 GHz, respectively. The position angles of the beam are $74.7\arcdeg$ and $70.6\arcdeg$. For continuum data, we used the 12-m array data only to drop spatially extended emission. We averaged the data over the bandwidths and cleaned with briggs weighting. The beam sizes and position angles are $1.71\arcsec\!\times\! 1.02\arcsec$, $73.2\arcdeg$, and $1.35\arcsec\!\times\! 0.55\arcsec$, $70.2\arcdeg$, at 230.5 GHz and 265.9 GHz, respectively.  

\subsection{N-body simulations}
Gravitational N-body simulations were performed to reproduce the distribution and kinematics of dense molecular clump adjacend to CO--0.40--0.22$^*$ by the gravitational kick model. We employed the leapfrog method to integrate Newton's equaion of motion. The time step was set to $4.6\times 10^3$ yr. We set a $10^5\,M_{\sun}$ point-like mass and a spherical cloud of a thousand 1 $M_{\sun}$ particles. The cloud is initially virialized, having a Gaussian radial dispersion of $0.2$ pc and velocity dispersion of $1.43$ \kms. The cloud was placed at $(X, Y)\!=\!(9.8\,\mbox{pc}, -0.65\,\mbox{pc})$, and then thrown with an initial velocity of $(v_{\rm X}, v_{\rm Y})\!=\!(-8.19\,\kms , 0.4\,\kms)$. The origin of the XY plane was set to the centre of gravity, which is almost coincident with the $10^5\,M_{\sun}$ point-like mass.  

Geometrical parameters were chosen basically in accordance with those of the previous paper\cite{Oka16}. The Y-axis was originally set parallel to the line of sight, then the XY plane was rotated about the $Z$-axis by $45\arcdeg$. The position angle of the orbital plane was set to $41.6\arcdeg$. We set the inclination ($i$) of the orbital plane to $70.0\arcdeg$ to reproduce the $\sim\!0.3$ pc displacement between the main HCN {\it J}=3--2 clump and CO--0.40--0.22$^*$.  The line of sight velocity of the centre of gravity was set to $\VLSR\!=\!-120$ \kms . The latter two parameters were $i\!=\!90.0\arcdeg$ and $\VLSR\!=\!-110$ \kms\ in the previous paper\cite{Oka16}.

\subsection{Data Availability}
The data that support the plots within this paper and other findings of this study are available from the corresponding author upon reasonable request. Data from our ALMA cycle 1 observations (ALMA/2012.1.00940.S) can be obtained from the ALMA Science Archive Query, http://almascience.nao.ac.jp/aq/.  

\end{methods}



\end{document}